# Clinical Decision Support Systems: A Visual Survey

Kamran Farooq[1], Bisma S Khan[2], Muaz A Niazi[2,*], Stephen J Leslie[3] and Amir Hussain[1]

[1]K. Farooq and A. Hussain are with Computing Science and Mathematics Division, University of Stirling, Scotland, UK (e-mail: kfa@cs.stir.ac.uk, ahu@cs.stir.ac.uk).
[2]B. S. Khan and M. A. Niazi are with the Department of Computer Science, COMSATS Institute of Information Technology, Islamabad, Pakistan (e-mail: bis.sarfraz@gmail.com, muaz.niazi@ieee.org).
[3] Raigmore Hospital, Inverness, Scotland, UK (e-mail: stephen.leslie@nhs.net).
*Corresponding author. E-mail: muaz.niazi@ieee.org

**Abstract:** Clinical Decision Support Systems (CDSS) form an important area of research. In spite of its importance, it is difficult for researchers to evaluate the domain primarily because of a considerable spread of relevant literature in interdisciplinary domains. Previous surveys of CDSS have examined the domain from the perspective of individual disciplines. However, to the best of our knowledge, no visual scientometric survey of CDSS has previously been conducted which provides a broader spectrum of the domain with a horizon covering multiple disciplines. While traditional systematic literature surveys focus on analyzing literature using arbitrary results, visual surveys allow for the analysis of domains by using complex network-based analytical models. In this paper, we present a detailed visual survey of CDSS literature using important papers selected from highly cited sources in the Thomson Reuters web of science. We analyze the entire set of relevant literature indexed in the Web of Science database. Our key results include the discovery of the articles which have served as key turning points in literature. Additionally, we have identified highly cited authors and the key country of origin of top publications. We also present the Universities with the strongest citation bursts. Finally, our network analysis has also identified the key journals and subject categories both in terms of centrality and frequency. It is our belief that this paper will thus serve as an important role for researchers as well as clinical practitioners interested in identifying key literature and resources in the domain of clinical decision support.

## 1. Introduction

The study of Clinical decision support systems (CDSS) constitutes a significant field of usage of information technology in healthcare. CDSS are designed to assist clinicians and other healthcare professionals in diagnosis as well as decision-making. CDSS uses healthcare data and a patient's medical history to make recommendations. By using a predefined set of rules, CDSS intelligently filters knowledge from complex data and presents at an appropriate time (Osheroff and Association 2006). By adopting CDSS, healthcare can become more accessible to large populations. However, it also implies that at times, CDSS may be used by people having literal medical knowledge (Ahn, Park et al. 2014).

Several researchers have contributed in the form of systematic literature reviews (SLR) and surveys to provide readers with insightful information about CDSS, as demonstrated below in Table 1.

**Table 1.** The existing literature review in the domain of clinical decision support systems

| Author | Ref. | Study Period | Survey Type | Study Area | Papers Reviewed |
|---|---|---|---|---|---|
| Ali et al. (2016) | (Ali, Giordano et al. 2016) | 2000-2014 | Systematic Review | Randomised control trials of CDSS | 38 |
| Vaghela et al. (2015) | (Vaghela, Bhatt et al. 2015) | 1987-2014 | Survey | Classification techniques of CDSS | 18 |
| Son et al. (2015) | (Son, Jeong et al. 2015) | 1979-2014 | Visualisation | E-Health | 3023 |
| Njie et al. (2015) | (Njie, Proia et al. 2015) | 1975-2012 | Systematic Review | CDSS and prevention of cardiovascular | 45 |

| | | | | diseases | |
|---|---|---|---|---|---|
| Madara (2015) | (Marasinghe 2015) | 1950-2014 | Systematic Review | CDSSs to improve medication safety in long-term care homes | 38 |
| Martínez-Pérez et al. (2014) | (Martínez-Pérez, de la Torre-Díez et al. 2014) | 2007-2013 | Literature and Commercial Review | Mobile CDSS and applications | 92 |
| Loya et al. (2014) | (Loya, Kawamoto et al. 2014) | 2004-2013 | Systematic Review | Service oriented architecture for CDSS | 44 |
| Fatima et al. (2014) | (Fathima, Peiris et al. 2014) | 2003-2013 | Systematic Review | CDSSs in the care asthma and COPD patients | 19 |
| Diaby et al. (2013) | (Diaby, Campbell et al. 2013) | 1960-2011 | Bibliometric | MCDA in healthcare | 2156 |
| Kawamoto et al. (2005) | (Kawamoto, Houlihan et al. 2005) | 1966-2003 | Systematic Review | Features of CDSS important for improving clinical practices | 70 |
| Chuang et al. (2000) | (Chuang, Hripcsak et al. 2000) | 1975-1998 | Methodological Review | Clustering in CDSS | 24 |

Despite the considerable variety of literature available, a key problem, researchers facing is the inability to understand the dynamics of CDSS-related literature. This is compounded due to the fact that this literature is spread across several related disciplines. Consequently, it is challenging to locate available information from a corpus of peer-reviewed articles. It is also difficult for researchers as well as clinical practitioners to comprehend the evolution of the research area.

SLR may easily get outdated, and may not meet specific requirements of a study, may not exist for new and emerging fields, and may be written for specific areas of interest. Whereas visual survey gives scientometric overview of the scientific literature, which provides a broader spectrum by embracing publications across multiple disciplines of the domain. Visual survey allows us to explore various trends and patterns in the bibliographic literature more efficiently and keeps our knowledge up to date.

In this paper, we present a visual survey of key literature from Web of Science (WoS) to provide a meaningful and valuable reference for further study in the field. We have used CiteSpace a key visually analytical tool for information visualization (Chen 2006). Although, CiteSpace has been used in a variety of disciplines, such as visual analysis of aggregation operator (Yu 2015), agent-based computing (Niazi and Hussain 2011), digital divide (Zhu, Yang et al. 2015), anticancer research (Xie 2015), tech mining (Madani 2015), and digital medicine (Fang 2015), etc. To the best of our knowledge, until now, there is no current review of recent literature on CDSS, which uses a scientometric analysis of networks formed from highly cited and important journal papers from the Web of Science (WoS).

The key contribution of this paper is the visual analysis of citations to give a scientometric overview of the diversity of the domain across its multiple sub-domains and the identification of core concepts. The ideas of visual analysisand survey stem from Cognitive Agent-based Computing framework [29] – a framework which allows for modeling and analysis of natural and artificial Complex Adaptive Systems.

In summary, the current paper identifies various important factors including the identification of the most important cluster in the cited references, visual analysis of the keyauthors, highly cited authors, key journals, core subject categories, countries of the origin of manuscripts, and the institutions. We hope that this work will assist researchers, academicians, and practitioners to learn about the key literature and developments in the CDSS domain.

The rest of CDSS survey is structured as: In Section II, we give a brief background of the CDSS. Next, in Section III, we present the methodology section including data collection and an overview of CiteSpace. This is followed by Section IV, containing results and discussion. Finally, Section VI concludes the paper.

## 2. Background

This section presents the necessary background of Decision Support System and CDSS.

### 2.1. Decision Support System (DSS)

The idea of DSS is very broad and different authors have defined it differently based on their research and roles DSS plays in the decision-making process (Druzdzel and Flynn 1999, Holsapple 2008). Some people regard DSS as a field of information management systems, whereas others consider it as an extension of management science systems (Keen 1980). Keen in his paper (Keen 1980) states that "there can be no definition of Decision Support Systems, only of Decision Support". Authors of (Finlay 1994) define it as "a computer-based system that aids the process of decision-making". Whereas the authors of (Turban 1990) define it as "an interactive, flexible, and adaptable computer-based information system, especially developed for supporting the solution of a non-structured management problem for improved decision-making. It utilises data, provides an easy-to-use interface, and allows for the decision maker's own insights." For further details, we encourage interested readers to see (Marakas , Ralph, Sprague et al. 1986, Silver 1991, Power 1997, Sauter 1997, Schroff 1998, Druzdzel and Flynn 1999, Power 2000, Power 2002).

#### 2.1.1. History

The notion of DSS has evolved in the late *1950*s, from the theoretical studies of organisational decision-making and in the early *1960*s from technical work on interactive computer systems (Keen and Scott 1978). The idea of assisting decision-makers using computers was published in *1963*(Bonini 1963). Scot Morton is known as one of the first researcher's groups who coined the term DSS (Scott 1971). Research on DSS has gained momentum in *1974*, and by *1979* nearly *30* case studies in the domain of DSS have been published (Keen 1980). Almost *271* applications of DSS have been published during the time span of May *1988* to *1994*(Eom, Lee et al. 1998).

#### 2.1.2. Architecture

Again, the architecture of DSS varies because different researchers have identified different components in DSS, e.g. (Sprague Jr and Carlson 1982, Haettenschwiler 2001, Power 2002). However, (Marakas) identifies five fundamental components of a generic DSS architecture: i) the user, ii) the data management system, iii) the knowledge engine, iv) the model management system, and v) the user interface.

#### 2.1.3. Classification

Once again, there is no universal classification of DSS; different researchers have proposed a different classification. Based on user criterion, authors classify as passive DSS, active DSS, and cooperative DSS (Haettenschwiler 2001). Whereas, based on the conceptual criterion, authors classify as data-driven, knowledge-driven, communication driven, model-driven DSS, and document-driven (Power 2002).

#### 2.1.4. Applications

DSS applications are adopted in several areas, such as business management (Bose and Sugumaran 1999), finance management (Zopounidis, Doumpos et al. 1997), forest management (Mendoza, Sprouse et al. 1991), medical diagnosis (Alickovic and Subasi 2016), wastemanagement (Bertanza, Baroni et al. 2016, Inglezakis, Ambăruş et al. 2016), oral anticoagulation management (Fitzmaurice, Hobbs et al. 1998), ship routing (Dong, Frangopol et al. 2016), ecosystem management (Rauscher 1999), value-based management (Hahn and Kuhn 2012), World Wide Web (Chen, Hong et al. 1999), diagnosis and grading of brain tumour(Tate, Underwood et al. 2006), and so on.

We intend to provide insight to CDSS researchers and practitioners about historical trends, current developments, and future directions of the CDSS domain.

### 2.2. Clinical Decision Support System

Since the beginning of computers, physicians and other healthcare professionals have expected the time when machines would aid them in the clinical decision-making and other restorative procedures. "CDSS provides clinicians, patients or individuals with knowledge and person-specific or population information, intelligently filtered or presented at appropriate times, to foster better health processes, better individual patient care, and better population health" (Osheroff and Association 2006). Ba and Wang use social network analysis in the domain of health-related online social neteorks (Ba and Wang 2013)

### 2.2.1. History

In the late 1950s, the very first articles regarding this provision appeared and within a few years, experimental prototypes were made available (Ledley and Lusted 1959). In 1970, three advisory systems have provided a useful overview of the origin of the work on CDSS: MYCIN system by Shortliffe for the selection of antibiotic therapy (Clancey, Shortliffe et al. 1979), a system by deDombal for the diagnosis of abdominal pain (Nugent, Warner et al. 1964, Clancey, Shortliffe et al. 1979), and a system called HELP for generating inpatient medical alerts (Warner 1979, Kuperman, Gardner et al. 2013).

### 2.2.2. Types

There exist two main types of CDSS. The first one is derived from expert systems and uses knowledge base. The knowledge base depends on inference engine to implement the rules, such as if-then-else on the patient data and presents the findings to end-users [2]. The second type of CDSS is based on the non-knowledge based systems, which depends on machine learning techniques for the analysis of clinical based data (Alther and Reddy 2015). The architectural parts in the conventional structures of CDSS consist of; user, knowledge base, inference engine and user interface (Bonney 2011).

### 2.2.3. Benefits

The key benefits of CDSS reported in the studies conducted in (Ivbijaro, Kolkiewicz et al. 2008, Haynes and Wilczynski 2010, Kawamoto, Del Fiol et al. 2010, Wright, Sittig et al. 2011, Musen, Middleton et al. 2014) are as follows:
1. Higher standards of patient safety: CDSSs have helped healthcare organisations all over the world acquiring higher standards of patient safety by adopting standardised clinical procedures governed by the clinical workflows encoded through these systems. Thus reducing diagnostic and prescribing errors and drug doubling issues.
2. Improving the quality of direct patient care: Their research also concluded that with the advent of CDSS, quality of care has improved to considerable levels with this extra support provided to clinicians (who are already struggling to cope with current healthcare demands). This has made it possible for clinical experts to allocate more time in providing direct patient care.
3. Standardisation and conformance of care using clinical practice guidelines: The standardisation of clinical pathways and procedures set precedents and evaluation benchmarks for healthcare trusts to achieve higher patient satisfaction levels set out by different healthcare organisations in different regions. CDSS also promote the utilisation of clinical practice guidelines (CPGs) for the development of knowledge-aware systems capable of performing effective clinical decision-making to promote standardised care.
4. Collaborative decision-making: CDSS have helped healthcare stakeholders that include clinicians, healthcare trusts and policy makers to develop safe and efficient care models using a collaborative decision-making approach to benefit both patient and a clinician. CDSS have also helped healthcare trusts to improve effectiveness in the prescribing facility through cost-effective drug order dispensation (Wright, Sittig et al. 2011). CDSS are also playing an important role in the integration of EHRs, which will help healthcare authorities to streamline information collection and clinical diagnosis operations in order to promote efficient data gathering (Ivbijaro, Kolkiewicz et al. 2008). The audit trail is another important aspect of modern healthcare systems which is achieved through the intelligent exploitation of clinical decision support capabilities.

### 2.2.4. Existing Reviews

Many reviews have identified the benefits of the CDSSs, in particular, Computerized Physician Order Entry systems (Hunt, Haynes et al. 1998, Eslami, de Keizer et al. 2008, Zuccotti, Maloney et al. 2014). The CDSS as part of the Computerized Physician Order Entry has been found to alleviate adverse drug events and medication errors (Jaspers, Smeulers et al. 2011, Steinman, Handler et al. 2011, Bright, Wong et al. 2012). CDSSs also have demonstrated to improve clinician performance, by way of promoting the electronic prescription of drugs, adherence to guidelines and to an extent the efficient use of time (Jaspers, Smeulers et al. 2011, Bright, Wong et al. 2012). CDSSs perform a key role in providing primary care and preventative measures at outpatient clinics, e.g. by alerting caregivers of the need for routine blood pressure checking, to recommend cervical screening, and to offer influenza vaccination (Hunt, Haynes et al. 1998, Ahmadian, van Engen-Verheul et al. 2011).

To provide effective healthcare delivery to patients, CDSS is used both in primary and secondary care units. In order to take maximum advantage from cardiovascular CDSS, it is required to ensure clinical governance in the next-generation clinical systems by considering a strong foundation in well-established clinical practice guidelines and evidence based medicine (Farooq and Hussain 2016).

### 2.2.5. CDSS Adoption

The adoption of CDSSs in diagnosis and management of chronic diseases, such as diabetes (O'Connor, Sperl-Hillen et al. 2011), cancer (Clauser, Wagner et al. 2011), dementia (Lindgren 2011), heart disease (DeBusk, Houston-Miller et al. 2010), and hypertension (Luitjes, Wouters et al. 2010) have played significant clinical roles in the main health care organisations in the improvement of clinical outcomes of the organisations worldwide at primary and secondary care. These CDSS also provide the foundation to system developer and knowledge expert to collate and build domain expert knowledge for screening by clinicians and clinical risk assessment (Khong and Ren 2011, Wright, Sittig et al. 2011).

An alternate approach to computer-assisted decision support was provided in the MYCIN development program, a clinical consultation system that de-emphasised diagnosis to concentrate on the appropriate management of patients who have infections (Shortliffe 1986).

### 2.2.6. Applications

CDSSs are considered as an important part in the modern units of healthcare organisations. They facilitate the patients, clinicians and healthcare stakeholders by providing patient-centric information and expert clinical knowledge (Classen, Phansalkar et al. 2011). To improve the efficiency and quality of healthcare, the clinical decision-making uses knowledge obtained from these smart clinical systems. The Automated DSSs of Cardiovascular are available in primary health care units and hospital in order to fulfil the ever-increasing clinical requirements of prognosis in the domain of coronary and cardiovascular diseases. The computer-based decision support strategies have already been implemented in various fields of cardiovascular care (Kuperman, Bobb et al. 2007). In the US and the UK, these applications are considered as the fundamental components of the clinical informatics infrastructures.

Ontology-driven DSS are being used widely in the clinical risk assessment of chronic diseases. The ontology-driven clinical decision support (CDS) framework for handling comorbidities in (Abidi, Cox et al. 2012) presented remarkable results in the disease management and risk assessment of breast cancer patients, which was deployed as a CDSS handling comorbidities in the healthcare setting for primary care clinicians in the Canada. They utilised semantic web techniques to model the clinical practice guidelines which were encoded in the form of a set of rules (through a domain-specific ontology) utilised by CDSSs for generating patient-specific recommendations.

Matt-MouleyBoumrane from the "University of Glasgow, 'UK" implemented an ontology-driven approach to the development of CDSS in the pre-operative risk assessment domain. In (Bouamrane, Rector et al. 2009), they reported their work by combining a preventative care software system in the pre-operative risk assessment domain with a decision support ontology developed with a logic based knowledge representation formalism. In (Farooq, Hussain et al. 2011, Farooq, Hussain et al. 2012, Farooq, Hussain et al. 2012), authors demonstrated utilisation of ontology and machine learning inspired techniques for the development of a hybrid CDS framework for cardiovascular preventative care. Their proposed CDS framework could be utilised for automatically conducting patient pre-visit interviews. Rather than replacing human experts, it would be used to prepare the patients before visiting a hospital, deliver educational materials, preorder appropriate tests, cardiac risk assessment scores, heart disease and cardiac chest pain scores. It would make better use of both patient and clinician time.

The ontology-driven recommendation and clinical risk assessment system could be used as a triage system in the cardiovascular preventative care which could help clinicians prioritize patient appointments after reviewing snapshot of patient's medical history (collected through an ontology-driven intelligent context-aware information collection using standardised clinical questionnaires) containing patient demographics information, cardiac risk scores, cardiac chest pain and heart disease risk scores, recommended lab tests and medication details. In (Farooq and Hussain 2016), they also have validated the proposed novel ontology and machine learning driven hybrid CDS framework in other application areas.

## 3. Methodology

In Figure 1, we illustrate the proposed methodology for the visual analysis of bibliographic literature in the domain of CDSS to uncover emerging patterns and trends.

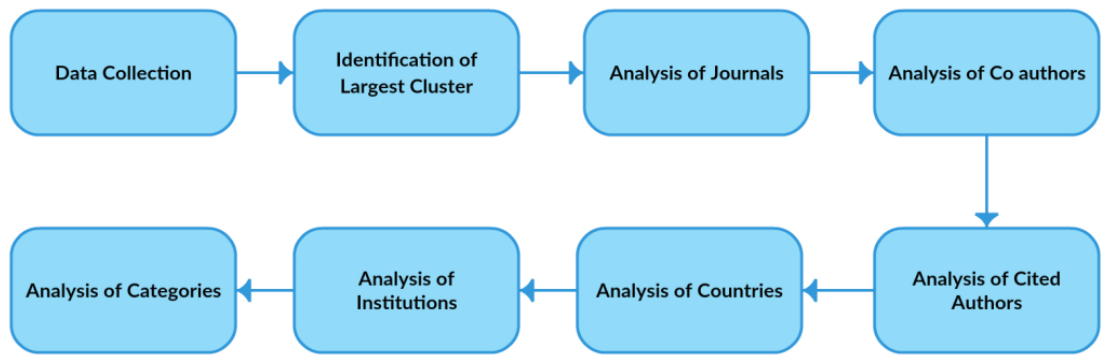

**Figure 1.** The proposed methodology (adapted from [2, 3]) for the visual analysis of clinical decision support system for the discovery of emerging patterns and trends in the bibliographic data of the domain.

### 3.1. Data Collection

The input dataset was collected from the Thomson Reuters Web of Science (Reuters 2008) between the timespan of *2005* to *2016*. Data was retrieved on *11 Nov 2016*, by an extended topic search for CDSSs including the web of science. The databases searched included SCI-Expanded, SSCI, and A&HCI. The search was confined to document types including articles, reviews, letters, and editorial material published in the English language. Each data record includes information as titles, authors, abstracts, and references. The input dataset contains a total of *1,945* records.

It is pertinent to note here that there is a problem in data collected from Web of Science. The WoS data identified two cited-authors named as "Anonymous" and "Institute of Medicine." In terms of frequency, Anonymous is the landmark node. However, on searching online it is found that WoS has picked it based on terms. Whereas on an extensive search on the internet, we found multiple papers having "Institute of Medicine" as an author.

### 3.2. CiteSpace: An Overview

In this research, we have used CiteSpace a key visually analytical tool for information visualisation(Chen 2006). CiteSpace is custom designed for visual analysis of citations. It uses colour coding to capture some details, which otherwise cannot be captured easily by using any other tool. In CiteSpace users can specify the years' range and the length of the time slice interval to build various networks. CiteSpace is based on network analysis and visualisation. It enables interactive visual analysis of a knowledge domain in different ways. By selecting display of visual attributes and different parameters, a network can be viewed in a variety of ways. CiteSpace has been used to analyse diverse domain areas such as agent-based computing (Niazi and Hussain 2011), cloud computing (Wu and Chen 2012), cross-language information retrieval (Rongying and Rui 2011), and clinical evidence (Chen and Chen 2005).

One of the key features of CiteSpace is the calculation of betweenness centrality (Chen 2006). The betweenness centrality score can be a useful indicator of showing how different clusters are connected (Chen 2016). In CiteSpace, the range of betweenness centrality scores is [0, 1]. Nodes which have high betweenness centrality are emphasised with purple trims. The thickness of the purple trims represents the strength of the betweenness centrality. The thicker the purple trim, the higher the betweenness centrality. A pink ring around the node indicates centrality >= *0.1*.

Burst identifies emergent interest in a domain exhibited by the surge of citations (Niazi and Hussain 2011). Citation bursts indicate the most active area of the research (Chen 2016). Burst nodes appear as a red circle around the node.

### 3.2.1. Colours Used

CiteSpace is designed for visualisation; it extensively relies on colours, therefore the description in this paper is based on colours.

The colours of the co-citation links personify the time slice of the study period of the first appearance of the co-citation link. Table 2demonstrates CiteSpace's use of colour to visualise time slices. Blue colour is for earliest years, the green colour is for the middle years, and orange and red colours are for the most recent years. A darker shade of the samecolour corresponds to earlier time-slice, whereas lighter shades correspond to the later time slice.

**Table 2**. Cite Space's use of colours to visualise links, and time slices.

| Link colours | Corresponding Time Slice |
| --- | --- |
| Blue | Earliest years |
| Green | Middle years |
| Orange and Redish | Most recent years |
| Darker shade of the same colour | Earliest time-slice |
| Lighter shade of the same colour | Later time-slice |

### 3.2.2. Node Types

The importance of a node can be identified easily by analysing the topological layout of the network. Three most common nodes, which are helpful in the identification of potentially important manuscripts are i) hub node, ii) landmark node, and iii) pivot node (Chen 2006).

Landmark nodes are the largest and most highly cited nodes. In CiteSpace, they are represented by concentric circles with largest radii. The concentric citation tree rings identify the citation history of an author. The colour of the citation ring represents citations in a single time slice. The thickness of a ring represents the number of citations in a particular time slice.

Hub nodes are the nodes with a large degree of co-citations.

Pivot nodes are links between different clusters in the networks from different time intervals. They are either gateway nodes or shared by two networks. Whereas turning points refer to the articles which domain experts have already identified as revolutionary articles in the domain. It is a node which connects different clusters by same coloured links.

## 4. Results and Discussion

This section briefly demonstrates results of our analysis.

### 4.1. Identification of the Largest Clusters in Document Co-Citation Network

To identify the most important areas of research, here we used cluster analysis. CiteSpace is used to form the clusters. It uses time slice to analyse the clusters. The merged network of cited references is partitioned into some major clusters of articles. In Figure 2, years from *2005* to *2016* show up as yearly slices represented by unique colours. We have selected top *50* cited references per one-year time slice. The links between the nodes also represent the particular time slices. In (Chen 2006) authors noted clusters with same colours are indicative of co-citations in any given time slice. The cluster labels start from 0; the largest cluster is labelled as (*#0*), the second largest is labelled as (*#1*), and so on. The largest cluster is the indicator of the major area of research.

It can also be noticed in the Fig. 2 that the articles of David W. Bates (1999) and Thomas D. Stamos (*2001*) are the intellectual turning points, which join two linked clusters: (cluster *#4*) "combination" and (cluster *#12*) "family practice" respectively. Similarly, articles of Heleen Van DerSijs (2008) and Blackford Middleton (*2013*) are the intellectual turning points, which join two linked clusters: (cluster *#2*) "decision support" and (cluster *#16*) "computerised prescriber order entry" respectively. After a gap of five years, Middleton B has cited a paper of Van DerSijs H, which drew the interest of many researchers in the field of "decision support".

It is interesting to note that the half-life of the article of Bates DW is *7* years and the half-life of the article of Thomas D. Stamos is *4* years. Whereas the half-life of Van Der Sijis H's article is *5* years and the half-life of Middleton B's paper is *3* years.

In Table 3, details of top five co-cited references are given in terms of high frequency. By observing this table, we observed that the top five articles have low centrality, but are still significant by having more frequency.

The article by Amit X. Garg (*2005)* has the highest frequency of citations among all the cited references. Following it are the articles of Kensaku Kawamoto and Gilad J. Kuperman published in *2005 and 2007* respectively. The articles of Van DerSijs H and Basit Chaudhry are also included in the top five articles of this domain.

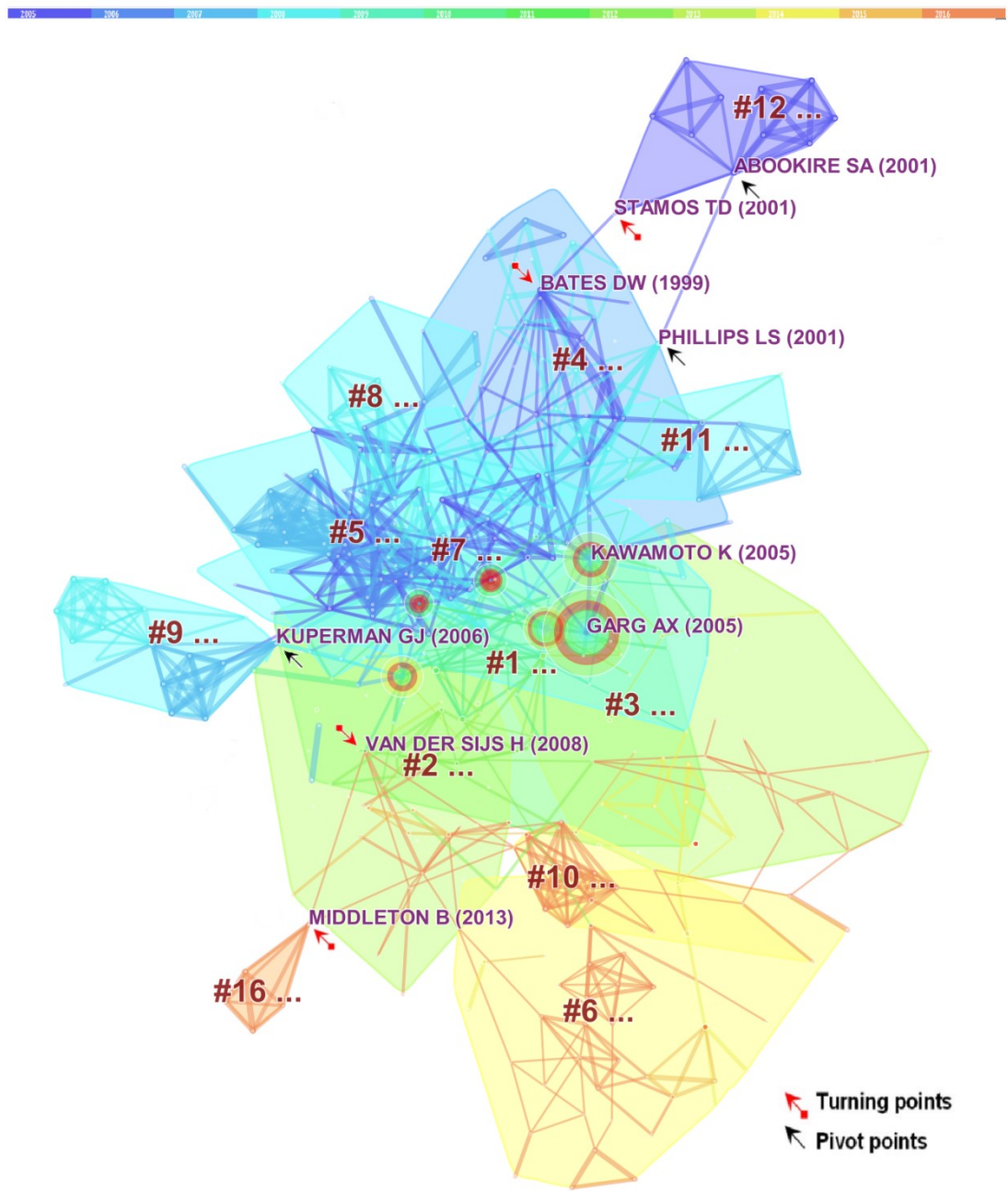

**Figure 2.** A merged network of cited references with *611* nodes and *1958* links on our CDSS dataset (*2005-2016*) based on *1*-year time slices. The largest component of connected clusters divided into *13* smaller clusters. The largest cluster(Niazi 2011) is "computerised decision support" and the smallest is "computerised prescriber order entry." The diameter of the circle corresponds to the frequency of the node. Whereas red circle indicates high citation burst of the article. The article of Garg AX has the highest frequency and highest citation burst among other articles of the domain.

Table 3 demonstrates documents in terms of frequency. It is also interesting to note that the article by Amit X. Garg (*2005*) is the landmark node with the largest radii. Amit X. Garg's article also has highest citation burst of *20.71,* which indicates that it has attracted huge attention from the research community. It has *223* citations *6*-year half-life, whereas it has*2357* citations on Google Scholar. Following it is the article of Kensaku Kawamoto (*2005*) with *15.46* citation burst, *151* citations, and half-life of *6* years. It has *1684* citations on Google Scholar. Next is the article by Kuperman GJ (*2007*) with *3.48* citation burst, *135*citation frequency, and a half-life of *5* years. It has *547* citations on Google Scholar. It is closely followed by the Van DerSijs H (*2007*) with a citation burst of *15.09*, citation frequency *116*, and half-life of *5* years. It has *690* citations on Google Scholar. The article byBasit Chaudhry (*2006*) has lowest citation burst of *2.99* among top five articles of the domain. It has a citation frequency of *112* and half-life of *6* years. It has *2491* citations on Google Scholar.

**Table 3.** The summary table of cited references sorted in terms of Frequency includes frequency (F), citation burst (CB), author (AU), publication year (PY), journal (J), Volume (V), page no. (PP), half-life (HL), cluster ID (CL), and Google Scholar Citations (GSC) of the top 5 most cited references.

| F | CB | AU | PY | J | V | PP | HL | CL | GSC |
|---|---|---|---|---|---|---|---|---|---|
| 223 | 20.71 | Garg AX | 2005 | JAMA-J AM MED ASSOC | V293 | P1223 | 6 | 3 | 2357 |
| 151 | 15.46 | Kawamoto K | 2005 | BRIT MED J | V330 | P765 | 6 | 3 | 1684 |
| 135 | 3.48 | Kuperman GJ | 2007 | J AM MED INFORM ASSN | V14 | P29 | 5 | 2 | 547 |
| 116 | 15.09 | Van der Sijs H | 2006 | J AM MED INFORM ASSN | V13 | P138 | 6 | 2 | 690 |
| 112 | 2.99 | Chaudhry B | 2006 | ANN INTERN MED | V144 | P742 | 5 | 1 | 2491 |

Table 4 contains cited documents in terms of betweenness centrality. The article by Basit Chaudhry (*2006*) is the most influential document with the highest centrality score of *0.43*. Half-life of this article is *5* years and it has *2491* citations on Google Scholar. Following it is the article by Ross Koppel (*2005*) with *0.24* centrality, and half-life of *5* years. It has *1995* citations on Google Scholar. Next is the article by Amit X. Garg (*2005*) with *0.18* betweennesscentraliy and a half-life of *6* years. It has *2357* citations on Google Scholar. It is closely followed by Jerome A. Osheroff (*2007*) with betweenness centrality of *0.16* and half-life of *5* years. It has *357* citations on Google Scholar. Finally, we have article by Gilad J. Kuperman (*2007*) with lowest betweenness centrality of *0.14* among top five articles of the domain. It has a half-life of *5* years. It has *547* citations on Google Scholar.

**Table 4.** The summary table of cited documents sorted in terms of Centrality includes betweenness centrality (BC), author (AU), publication year (PY), journal (J), Volume (V), page no. (PP), half-life (HL), cluster ID (CL), and Google Scholar Citations (GSC) of the top 5 most cited references.

| BC | AU | PY | J | V | PP | HL | CL | GSC |
|---|---|---|---|---|---|---|---|---|
| 0.43 | Chaudhry B | 2006 | ANN INTERN MED | V144 | P742 | 5 | 4 | 2491 |
| 0.24 | Koppel R | 2005 | JAMA-J AM MED ASSOC | V293 | P1197 | 5 | 0 | 1995 |
| 0.18 | Garg AX | 2005 | JAMA-J AM MED ASSOC | V293 | P1223 | 6 | 4 | 2357 |
| 0.16 | Osheroff JA | 2007 | J AM MED INFORM ASSN | V14 | P141 | 5 | 4 | 357 |
| 0.14 | Kuperman GJ | 2007 | J AM MED INFORM ASSN | V14 | P29 | 5 | 1 | 547 |

The merged network contains a total of *611* cited references and *1,958* co-citation links. The largest cluster, i.e. (*#0*) of the network is disconnected from the largest component of the network. In this analysis, we will consider only largest component.

The largest component of connected clusters contains *442* nodes, which is *72%* of the network. The largest component is further divided into *13* smaller clusters of different sizes. Table 5 illustrates the details of these clusters.

Cluster *#1* (largest cluster) contains *65* nodes, which are *10.628%* of whole nodes in the network. The average publication year of the literature in this cluster is *2007*. The mean silhouette score of *0.737* indicates relatively high homogeneity in the cluster.

Cluster *#2* contains *57* nodes, which are *9.328%* of whole nodes in the network. The average publication year of the literature in this cluster is *2009*. The mean silhouette score of *0.7* indicates relatively high homogeneity in the cluster.

Cluster *#3* contains *56* nodes, which are *9.165%* of whole nodes in the network. The average publication year of the literature in this cluster is *2008*. The mean silhouette score of *0.722* indicates relatively high homogeneity in the cluster. It is interesting to note that cluster *#3*("AIDS") contains several articles with strongest citation burst, which indicates it is an active or emerging area of research.

Cluster *#4* contains *52* nodes, which are *8.51%* of whole nodes in the network. The average publication year of the literature in this cluster is *2001*. The mean silhouette score of 0.791 indicates average homogeneity in the cluster. It is interesting to note that most of the highly influential articles are the members of cluster *#4*.

Cluster *#5* contains *49* nodes, which are *8.01%* of whole nodes in the network. The average publication year of the literature in this cluster is *2003*. The mean silhouette score of 0.772 indicates relatively high homogeneity in the cluster.

Cluster *#6* contains *45* nodes, which are *7.364%* of whole nodes in the network. The average publication year of the literature in this cluster is *2012*. The mean silhouette score of *0.955* indicates very high homogeneity in the cluster.

Cluster *#7* contains *40* nodes, which are *6.546%* of whole nodes in the network. The average publication year of the literature in this cluster is *2002*. The mean silhouette score of *0.73* indicates relatively high homogeneity in the cluster.

Cluster *#8* contains 1*9* nodes, which are *3.10%* of whole nodes in the network. The average publication year of the literature in this cluster is *2003*. The mean silhouette score of *0.854* indicates high homogeneity in the cluster.

Cluster *#8* contains *19* nodes, which are *3.10%* of whole nodes in the network. The average publication year of the literature in this cluster is *2003*. The mean silhouette score of *0.854* indicates high homogeneity in the cluster.

Cluster *#9* contains *18* nodes, which are *2.945%* of whole nodes in the network. The average publication year of the literature in this cluster is *2004*. The mean silhouette score of *0.976* indicates very high homogeneity in the cluster.

Cluster *#10* contains *13* nodes, which are *2.127%* of whole nodes in the network. The average publication year of the literature in this cluster is *2011*. The mean silhouette score of *0.976* indicates very high homogeneity in the cluster.

Cluster *#11* contains *12* nodes, which are *1.963%* of whole nodes in the network. The average publication year of the literature in this cluster is *2002*. The mean silhouette score of *0.944* indicates very high homogeneity in the cluster.

Cluster *#12* contains *11* nodes, which are *1.800%* of whole nodes in the network. The average publication year of the literature in this cluster is *1999*. The mean silhouette score of *0.979* indicates very high homogeneity in the cluster.

Cluster *#16* (smallest cluster) contains *5* nodes, which are *0.818%* of whole nodes in the network. The average publication year of the literature in this cluster is *2010*. The mean silhouette score of *0.955*indicates very high homogeneity in the cluster.

**Table 5.** The summary table of Largest clusters of the co-cited authors. It contains the ID of the cluster, the size of the cluster, the average publication year of the literature in the cluster, and title terms of the clusters. The merged network contains 611 nodes and 1958 connections.

| Cluster ID | Size | Silhouette | Mean (Year) | Label (Log-Likelihood Ratio) | Terms (Mutual Information) |
|---|---|---|---|---|---|
| 1 | 65 (10.638%) | 0.737 | 2007 | Impact; Adverse Drug Event; Physician Order Entry; | Computerized Decision Support |
| 2 | 57 (9.328%) | 0.7 | 2009 | Alert; Ambulatory Care; Safety Alert; | Drug Administration |
| 3 | 56 (9.165%) | 0.722 | 2008 | Patient Outcome; Management; Guideline; | Aid |
| 4 | 52 (8.51%) | 0.791 | 2001 | Decision Support System; Primary Care; Expert System; | Combination |
| 5 | 49 (8.01%) | 0.772 | 2003 | Adverse Drug Event; Medication Error; Prevention; | Chronic Illness |
| 6 | 45 (7.364%) | 0.955 | 2012 | Personalized Medicine; Pharmacogenomics; Computed Tomography; | ACR Appropriateness Criteria |
| 7 | 40 (6.546%) | 0.73 | 2002 | Prevention; Intervention; Adverse Drug Event; | Acute Kidney Failure |
| 8 | 19 (3.10%) | 0.854 | 2003 | Emergency Medicine; ASHP; Systems Analysis; | Intra Cluster Correlation Coefficient |
| 9 | 18 (2.945%) | 0.976 | 2004 | Personal Digital Assistant; Resource; PDA; | Consultation |
| 10 | 13 (2.127%) | 0.976 | 2011 | Medication Alert System; Interview; Observational Study; | Surgery |
| 11 | 12 (1.963%) | 0.944 | 2002 | Guideline Implementation; Adverse Event; Clinical Practice Guideline; | Factor-V-Leiden |
| 12 | 11 (1.800%) | 0.979 | 1999 | Statin; Cholesterol Reduction; Treatment Panel III; | Family Practice |
| 16 | 5 (0.818%) | 0.995 | 2010 | Smoking Cessation; Control Intervention; Usability; | Computerized Prescriber Order Entry |

After an overview of the identification of clusters in the cited reference network, next, we move to the analysis of the journals.

**4.2. Analysis of Journals**

In this section, we visualise cited journals. Out of *1,945* records in the dataset, the *60* most cited journals were selected per one-year slice to build the network.

The pink rings around the nodes depicted in Figure 3**Error! Reference source not found.** indicate that there are five nodes in the network with centrality >*0.1*. "Journal of the American Medical Informatics Association" has the largest number of highly cited publications. The second largest number of publications is associated with the "The Journal of the American Medical Association." "Proceedings of the AMIA Symposium" (*2005*) has the strongest citation burst among authors from the period of *2005*.

Table 6 gives details of the top *5* key journals based on centrality. "The Journal of the American Medical Association" has the highest centrality score of *0.14* among all the other journals. It has *37.684* impact factor. In addition, it could be seen that in terms of centrality, the "Journal of the American Medical Informatics Association," the "International Journal of Medical Informatics," "The American Journal of Medicine" and the "Artificial Intelligence in Medicine" are also some of the productive journals of this domain with a centrality score of *0.13* and impact factor of *3.428*, *2.363*, *5.610*, and *2.142* respectively.

**Table 6.** In terms of centrality, the five most productive journals in the bibliographic literature of the CDSS domain. Jama-j AM MED ASSOC is the most central journal with centrality score 0.14, whereas ArtifIntell Med is the least central journal with centrality score 0.13.

| Centrality | Title | Abbreviated Title | Impact Factor |
|---|---|---|---|
| 0.14 | The Journal of the American Medical Association | Jama-j AM MED ASSOC | 37.684 |
| 0.13 | Journal of the American Medical Informatics Association | J AM MED INFORM ASSN | 3.428 |
| 0.13 | International Journal of Medical Informatics | Int J MED INFORM | 2.363 |
| 0.13 | The American Journal of Medicine | Am J MED | 5.610 |
| 0.13 | Artificial Intelligence in Medicine (AIIM) | Artif INTELL MED | 2.142 |

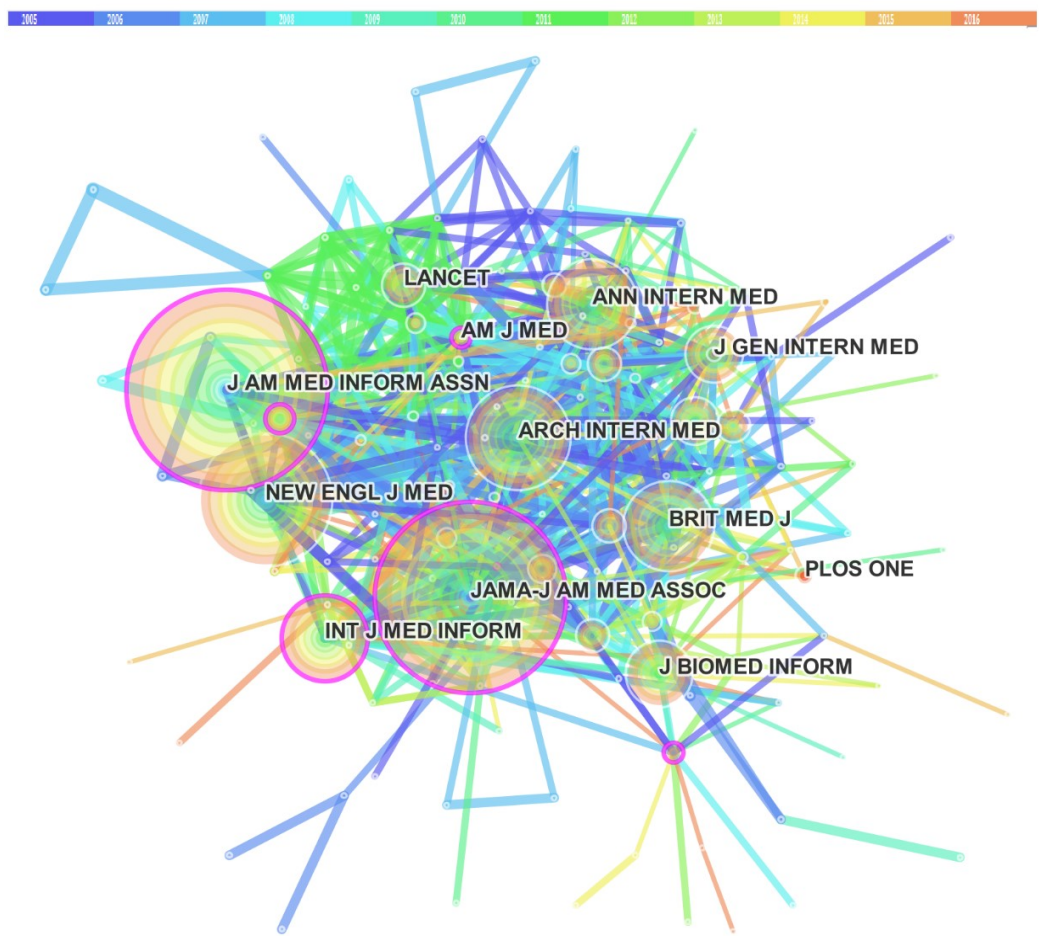

**Figure 3.** Journals' network in terms of centrality. Concentric citation tree rings indicate the citation history of the publications of a journal. The colours of the circles in the tree ring represent citations in a corresponding year. The red rings indicate the citation burst of the publication. The colours of the links correspond to the time slice. The pink rings around the node indicate the centrality >= 0.1. The "J AM MED INFORM ASSN" is the highly cited journal, whereas the "Jama-j AM MED ASSOC" is the most central Journal of the domain

Table 7 gives details of the top *5* key journals based on their frequency of publications. It is interesting to note that the table organised in terms of frequency of publication gives a somewhat different set of key journals. The "Journal of the American Medical Informatics Association" is at the top with a frequency of *1169* publications and *3.428* impact factor. This is followed by "The Journal of the American Medical Association", "The New England Journal of Medicine," "The Archives of Internal Medicine", and the "Annals of Internal Medicine Journal" with frequencies *1961*, *819*, *687*, and *655* and impact factor *37.684, 59.558, 17.333*, and *16.593* respectively.

**Table 7.** The five most productive journals in the bibliographic literature of the CDSS domain based on frequency. J AM MED INFORM ASSN is the most cited journal with frequency 1169, whereas Ann INTERN MED is the least cited journal with frequency 655.

| Frequency | Title | Abbreviated Title | Impact Factor (2016) |
|---|---|---|---|
| 1169 | Journal of the American Medical Informatics Association (JAMIA) | J AM MED INFORM ASSN | 3.428 |
| 1096 | The Journal of the American Medical Association | Jama-j AM MED ASSOC | 37.684 |
| 819 | The New England Journal of Medicine (NEJM) | New ENGL J MED | 59.558 |
| 687 | Archives of Internal Medicine | Arch INTERN MED | 17.333 |
| 655 | Annals of Internal Medicine Journal | Ann INTERN MED | 16.593 |

After a visual analysis of the journals, in the next section, we will analyse the authors' network.

**4.3. Analysis of Co-Authors**

This section analyses the author collaborative network. Figure 4 displays the visualisation of the core authors of the domain. The merged network contains *346* authors and *719* co-citation links. As shown in Fig. 4, burst nodes appear as a red circle around the node. The citation burst in authors network specifies the authors who have rapidly grown the number of publications.

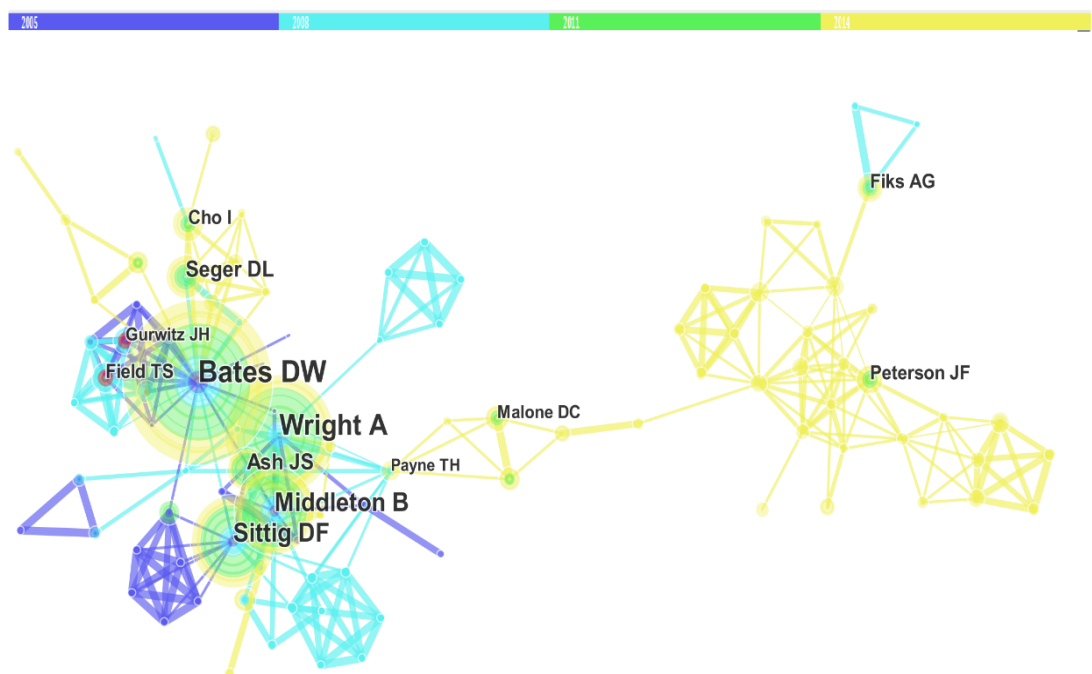

**Figure 4.** Co-authors network visualisation. The merged network contains *346* nodes and *719* links. Top 20% nodes are selected per slice (of length 3). Burst nodes appear as a red circle around the node. Concentric citation tree rings indicate the citation history of the publications of an author. David BW is the highly cited node with the frequency of 59, whereas Payne TH is the most central node with a centrality score of 0.08. Gurwitz JH and Field TS have longest citation burst periods.

As shown in Figure 5, in terms of frequency, David BW is the landmark node with largest radii of the citation ring. Payne TH is the most central author of this domain.

Visualisation in Figure 5 illustrates the authors who have the strongest citation bursts and years in which it took place. It can be seen that Ali S. Raja (*2014*) from "Harvard Medical School, USA" has the strongest burst among the top *5* authors since *2005*. Ivan K. Ip (*2005*) from "Harvard Medical School, USA" has the second strongest burst, which took place in the period of *2013* to *2016*. Following him are Terry S. Field (*2005*) from Meyers Primary Care Institute, Ramin Khorasani (*2014*) from "Brigham and Women's Hospital", and Jerry H. Gurwitz (*2005*) from "Meyers Primary Care Institute, USA."

| Authors | Year | Strength | Begin | End | 2005 – 2016 |
|---|---|---|---|---|---|
| Raja AS | 2005 | 5.2538 | 2014 | 2016 | |
| Ip IK | 2005 | 4.4697 | 2013 | 2016 | |
| Field TS | 2005 | 4.4586 | 2005 | 2009 | |
| Khorasani R | 2005 | 4.3956 | 2014 | 2016 | |
| Gurwitz JH | 2005 | 4.0089 | 2005 | 2009 | |

**Figure 5.** The top 5 Co-authors associated with strongest citation bursts. The history of the burstness of authors includes names of the authors, publication year, burst strength, starting and ending year of the citation burst. "Raja AS" has strongest citation burst among all other authors. "Field TS" and "Gurwitz JH" have the longest burst period.

Even though this visualisation gives a general picture of the several authors, Table 8 also illustrates a comprehensive analysis of authors' network. Here we can notice that highly cited author in the network is David Bates with *59* citations. David Bates is a Prof. of Medicine at "Harvard Medical School, USA." His areas of interest are medication safety, patient safety, quality, medical informatics, and clinical decision support. Next is Adam Wright, an Assoc. Prof. of Medicine, "Harvard Medical School, USA" and "Brigham and Women's Hospital, USA". His areas of interest are health information technology, medical informatics, biomedical informatics, clinical information systems, and CDS. Dean F. Sittig is the Cristopher Sarofilm Family Prof. of Bioengineering, "Biomedical Informatics, and UTHealth, USA." CDS, electronic health records, medical informatics, and biomedical informatics are his areas of interest. Next is Blackford Middleton, an Instructor, "Harvard TH Chan School of Public Health, USA". His areas of interest include personal health record, clinical informatics, CDs, knowledge management, and electronic medical record. Finally, we have RaminKhorasani, MD, PhD, "Brigham and Women's Hospital, USA."

**Table 8.** The top 5 Authors in terms of the frequency. David Bates is the most cited author with 395 citations.

| Frequency | Author | Abbreviations |
|---|---|---|
| 395 | David Bates | BATES DW |
| 296 | Amit X. Garg | GARG AX |
| 255 | Kensaku Kawamoto | KAWAMOTO K |
| 180 | Rainu Kaushal | KAUSHAL R |
| 173 | Gilad J. Kuperman | KUPERMAN GJ |

For additional relative analysis, we have observed the top-cited authors based on centrality, as depicted in Table 9. Thomas Payne a Prof. of Medicine, "University of Washington, USA." His areas of interest are clinical informatics and clinical computing. Richard D Boyce, Asst. Prof. of "Biomedical Informatics, University of Pittsburgh, USA". His areas of interest are Pharmacoepidemiology, medication safety, knowledge representation, comparative effectiveness research, and semantic web. Next is Robert R Freimuth, "Mayo Clinic, USA." His areas of interest include genomics CDS, personalised medicine, genetic variation, data integration, Pharmacogenomics, data integration and interoperable infrastructure. Matthias Samwald, "Medical University of Vienna, Austria." His interest is in biomedical informatics.

**Table 9.** The top 5 Co-Authors in terms of centrality. Payne TH is the most central author with a centrality score 0.08, whereas the rest of the authors have the same centrality score 0.07.

| Centrality | Author | Abbreviations | Year |
|---|---|---|---|
| 0.08 | Thomas Payne | Payne TH | 2008 |
| 0.07 | David Bates | Bates DW | 2005 |
| 0.07 | Richard D Boyce | Boyce RD | 2014 |
| 0.07 | Robert R Freimuth | Freimuth RR | 2014 |
| 0.07 | Matthias Samwald | Samwald M | 2014 |

After analysing authors' network, in the next section, we have visualised the cited authors' network.

### 4.4. Analysis of Cited-Authors

This section analyses the authors' co-citation network. Figure 6 displays the visualisation of the cited authors of this domain. The merged network contains *211* cited authors and *656* links. Burst nodes appear as a red circle around the node; the citation burst in cited-authors network specifies the authors who have rapidly grown the number of citations. In terms of frequency, David BW is the landmark node with largest radii of the citation ring. The pink ring around David BW indicates that it is also the most central author of this domain.

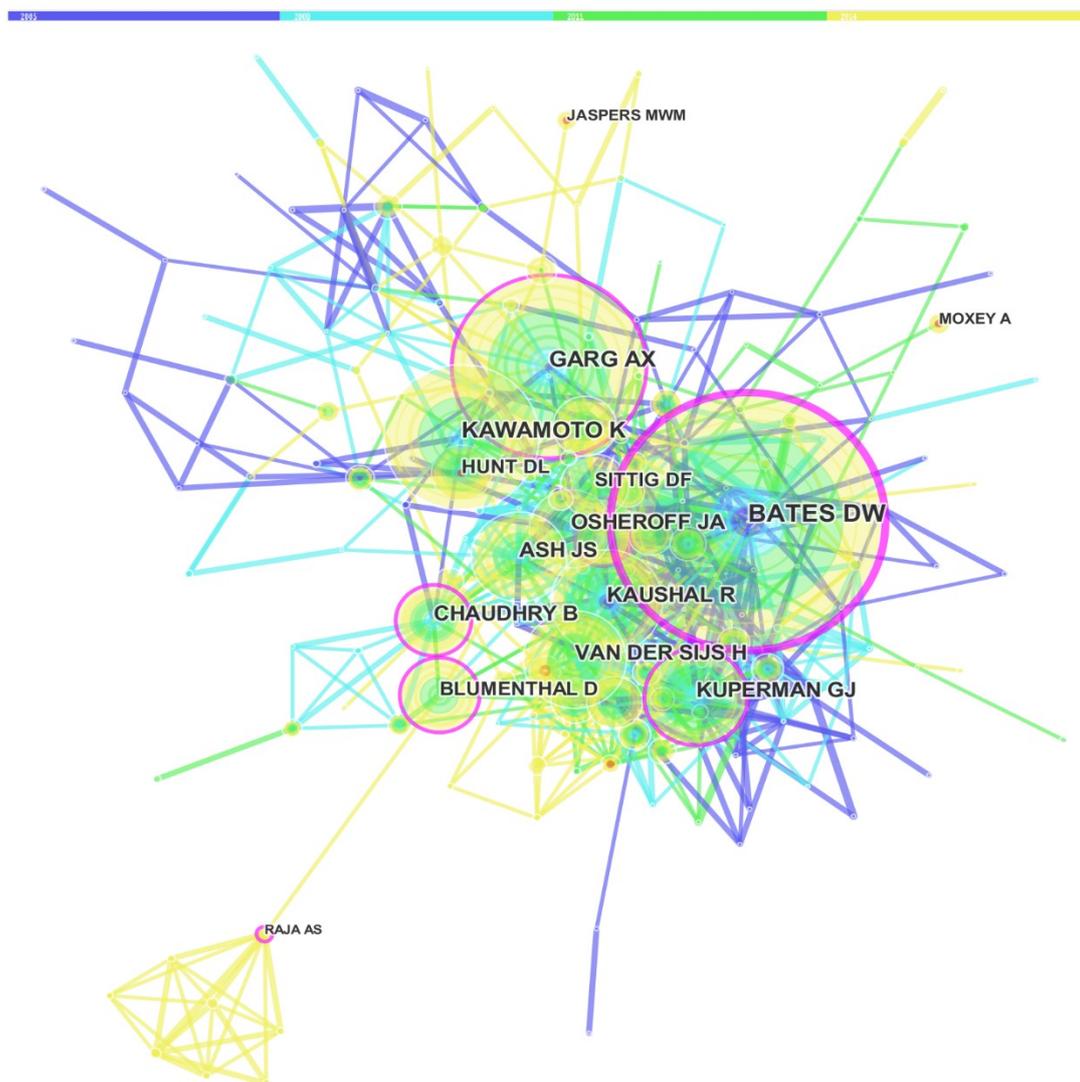

**Figure 6.** Cited-authors network visualisation. The merged network contains *211* nodes and *656* links. Burst nodes appear as a red circle around the node. Concentric citation tree rings indicate the citation history of the publications of an author. The pink rings around the node indicate the centrality score >= *0.1*. Bates DW is the landmark with largest radii and is also the hub node with the highest degree.

Even though this visualisation gives a general picture of the several authors, Table 10 also illustrates a comprehensive analysis of authors' network. Here we can notice that highly cited author in the network is David Bates with *460* citations. Next is Amit X. Garg, a Prof. of Medicine (Nephrology), Biostatistics & Epidemiology, "Western University, Canada". His areas of interest are kidney diseases, kidney donation, and clinical research. Following him is Kensaku Kawamoto, an Asst. Prof. of Biomedical Informatics and Assoc. CMIO in the "University of Utah, USA". Knowledge management, CDS, and standards and interoperability are his areas of interest. Next is Rainu Kaushal, "Departments of Medicine, Quality Improvement, Risk Management, and Children's Hospital, Boston, Massachusetts, USA". Finally, we have Gilad J. Kuperman, an Adjunct Assoc. Prof. of Biomedical Informatics, "Columbia University Clinical Informatics, USA".

**Table 10.** The Top 5 Cited-Authors in terms of the frequency. David Bates is the most cited author with 460 citations, whereas Kuperman GJ is the least cited author with 198 citations.

| Frequency | Author | Abbreviations | Year |
|---|---|---|---|
| 460 | David Bates | Bates DW | 2005 |
| 338 | Amit X. Garg | Garg AX | 2005 |

| 280 | Kensaku Kawamoto | Kawamoto K | 2005 |
| 207 | Rainu Kaushal | Kaushal R | 2005 |
| 198 | Gilad J. Kuperman | Kuperman GJ | 2005 |

For additional comparative analysis, we have observed the top-cited authors in terms of centrality. Fresh names which enter in Table 11are: David Blumenthal from the "Harvard Medical School, USA" and Basit Chaudhry from the "University of California, USA."

**Table 11.** The top 5 Cited-Authors in terms of centrality. Bates DW is the most central author with a centrality score of *0.29*, whereas Chaudhry B is the least central author with a centrality score of 0.12.

| Centrality | Author | Abbreviations | Year |
|---|---|---|---|
| 0.29 | David Bates | Bates DW | 2005 |
| 0.13 | Gilad J. Kuperman | Kuperman GJ | 2005 |
| 0.13 | Amit X. Garg | Garg AX | 2005 |
| 0.13 | David Blumenthal | Blumenthal D | 2009 |
| 0.12 | Basit Chaudhry | Chaudhry B | 2007 |

After analysing authors' network, in the next section, we will visualise the countries of the origin of the key publications of the domain.

### 4.5. Analysis of Countries

In this section, we demonstrate a visual analysis of the spread of research in the domain from different countries**.** For this visualisation, top *30* countries are chosen from the entire time span of *16* years (i.e. *2005-2016*) for each one-year time slice. In Fig. 7, the concentric rings of different colours represent papers published in different time slices. The diameter of the ring thus indicates the frequency of the country. From the display, it can be seen that the "United States" has the highest frequency, which indicates that the origin of key publications in the domain is the "United States". This is followed by articles originating from England, Canada, Netherlands, and Australia. The pink circle around the node represents the centrality >= *0.1*. As depicted inFigure 7, the Canada has highest centrality value. This is followed by the US, England, Germany, and Spain. Red circles represent the citation burst. The Scotland has the strongest citation burst, which provides the evidence that the articles originating in the domain from the Scotland have attracted a degree of attention from its research community.

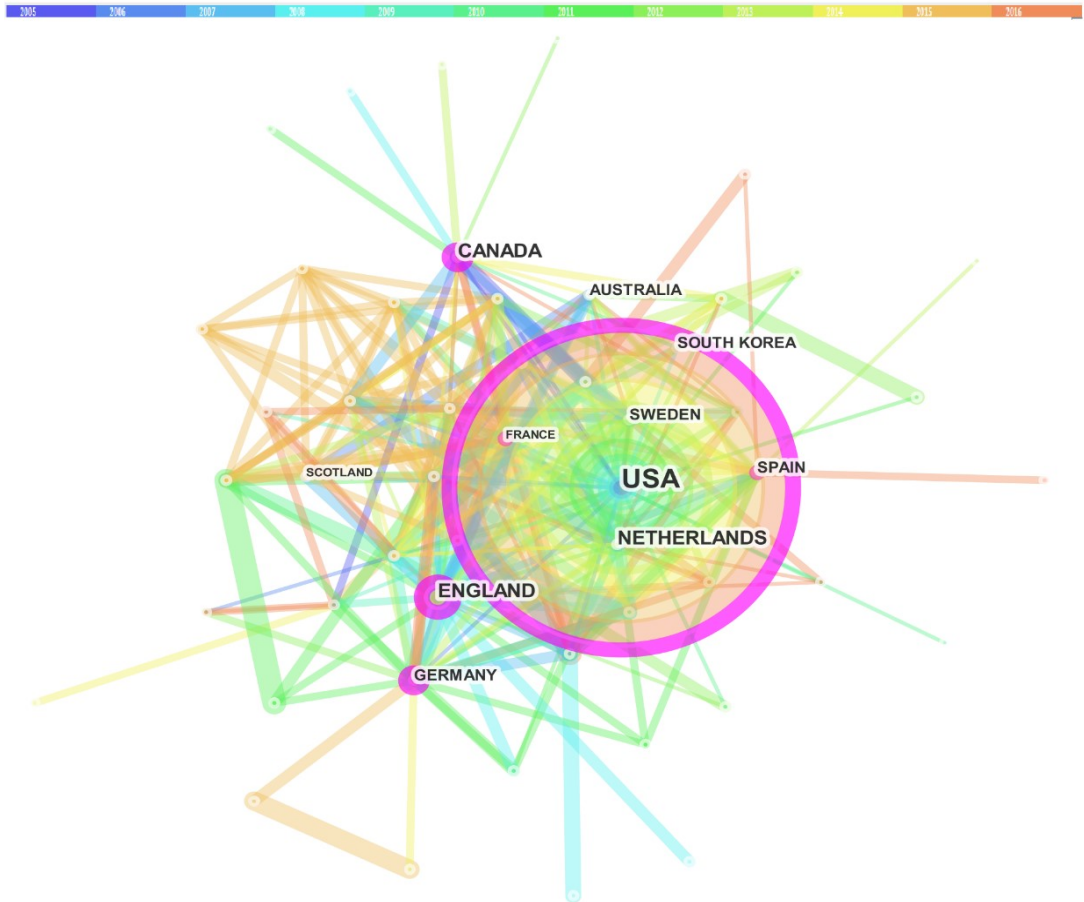

**Figure 7.** Countries network of *55* nodes and *263* links. The burst nodes appear as a red circle around the node. Concentric citation tree rings indicate the citation history of the publications of a country. The pink circle around the node represents the centrality >= *0*. The USA is the highly cited node, whereas Canada is the most central node and Scotland has strongest citation burst.

After a visual analysis of countries, we will present a visual analysis of institutions of highly cited publications.

### 4.6. Analysis of Institutions

In this section, visualisation of institutions is performed. Figure 8 contains a merged network of institutions of *319* nodes and *844* edges. We have selected top *50* nodes per one-year length time slice from *1,945* records. The "Harvard" is the most central, as well as highly cited node among all other institutions. Following it is the "Brigham and Women's Hospital, USA." Whereas the "University of Massachusetts, USA" has the strongest citation burst.

**Figure 8.** The network of Institutions, containing *319* nodes and *844* edges. Concentric citation tree rings demonstrate the citation history of the publications of an institution. The purple circle represents betweenness centrality. The thicker the purple ring, the higher the centrality score. The "University of Massachusetts" has the strongest burst. The Harvard is the highly cited and most central institution of the domain.

A visual analysis of the history of the burstness of institutions identifies universities that are specifically active in the research in this domain. As shown in Figure 9, the "University of Massachusetts, USA" has the strongest and longest citation burst among all other institutes in the timespan of *2006* to *2009*. The "Indiana University School of Medicine, USA" also has the longest period of the burst from *2013* till *2016*. Whereas, the "Weill Cornell Graduate School of Medical Sciences, USA" has shortest citation burst.

| Institutions | Year | Strength | Begin | End | 2005 - 2016 |
|---|---|---|---|---|---|
| Univ Massachusetts | 2005 | 6.2883 | 2006 | 2009 | |
| Indiana Univ Sch Med | 2005 | 6.1261 | 2013 | 2016 | |
| Cleveland Clin Fdn | 2005 | 4.1361 | 2010 | 2011 | |
| Johns Hopkins Univ | 2005 | 4.0639 | 2012 | 2013 | |
| Weill Cornell Med Coll | 2005 | 3.9329 | 2011 | 2012 | |

**Figure 9.** History of the burstness of institutions includes names of institutions, year of publication, the strength of burstness, beginning and ending year of the citation burst. The "University of Massachusetts" has the strongest burst, whereas the "University of Massachusetts" and the "Indiana University School" have the longest period of burst among all other institutions.

Next, we performed an analysis in terms of the frequency of publications associated with the institutions. Table 12 represents the top five institutions based on frequency. The "Havard, USA" has the highest ranking with the frequency of *165* publications. The "Brigham & Women's Hospital, USA" followed it closely with the frequency of *122* publications. Next is the "Vanderbilt University, USA" with the frequency of *62* publications. With *56* publications, next, we have the "University of Utah, USA". Following it, we have the "University of Washington, USA" with the frequency of *55* publications.

Table 12. The top Institutions in terms of Frequency. "Harvard" has the highest frequency of *165*, whereas the "University of Washington" has the lowest frequency of *55*.

| Frequency | Institution | Countries |
|---|---|---|
| 165 | Harvard University | USA |
| 122 | Brigham and Women's Hospital | USA |
| 62 | Vanderbilt University | USA |
| 56 | University of Utah | USA |
| 55 | University of Washington | USA |

In the Table 13 below, we performed another analysis in terms of the centrality of the publications. Table 13contains the list of the top five universities based on the centrality. It is interesting to note that top two universities the "Harvard" and "Brigham & Women's Hospital, USA" with centrality scores *0.3* and *0.17* respectively are also the highly cited institutions. Following them is the "University of Utah, USA" with centrality score *0.14*. Next is the "University of Washington, USA" with centrality score *0.09*. With centrality value *0.07*, it seems however that the "Heidelberg University, USA" has the lowest centrality score among all other institutions.

Table 13. The top 5 institutions in terms of the centrality. Top most University has a centrality score of *0.3*, whereas the "Heidelberg University" has a lowest centrality score of *0.07*.

| Centrality | Institutions | Countries |
|---|---|---|
| 0.3 | Harvard University | USA |
| 0.17 | Brigham and Women's Hospital | USA |
| 0.14 | University of Utah | USA |
| 0.09 | University of Washington | USA |
| 0.07 | Heidelberg University | Germany |

After visualisation of institutions, in the next section, we will present an analysis of subject categories of the domain.

### 4.7. Analysis of Categories

In this section, our next analysis is to discover publications associated with various categories. Fig. 10 depicts the temporal visualisation of categories in the domain.This merged network contains *95* categories and *355* links (co-occurences). We have selected top *50* nodes per one-year time slice. The detailed analysis based on the centrality and frequency is given below.

Table 14lists the top *5* categories based on centrality.The category "Health Care Sciences & Services" leads over other categories with centrality value *0.29*. It is closely followed by "Engineering" with centrality *0.28*. Next is "Computer Science" with centrality score *0.25*. Following it is the "Surgery" with centrality *0.18*. Subsequently, we have "Nursing" with centrality score *0.24*.

For relative analysis, we have also analysed these categories in terms of frequency of publications of manuscripts. The outcomes of this analysis are illustrated underneath inTable 15.

Table 14. The top 5 categories based on centrality. The subject category "HEALTH CARE SCIENCES & SERVICES" leads over other categories with a centrality score of 0.29.

| Centrality | Category |
|---|---|
| 0.29 | Health Care Sciences and Services |
| 0.28 | Engineering |
| 0.25 | Computer Science |
| 0.18 | Surgery |
| 0.16 | Nursing |

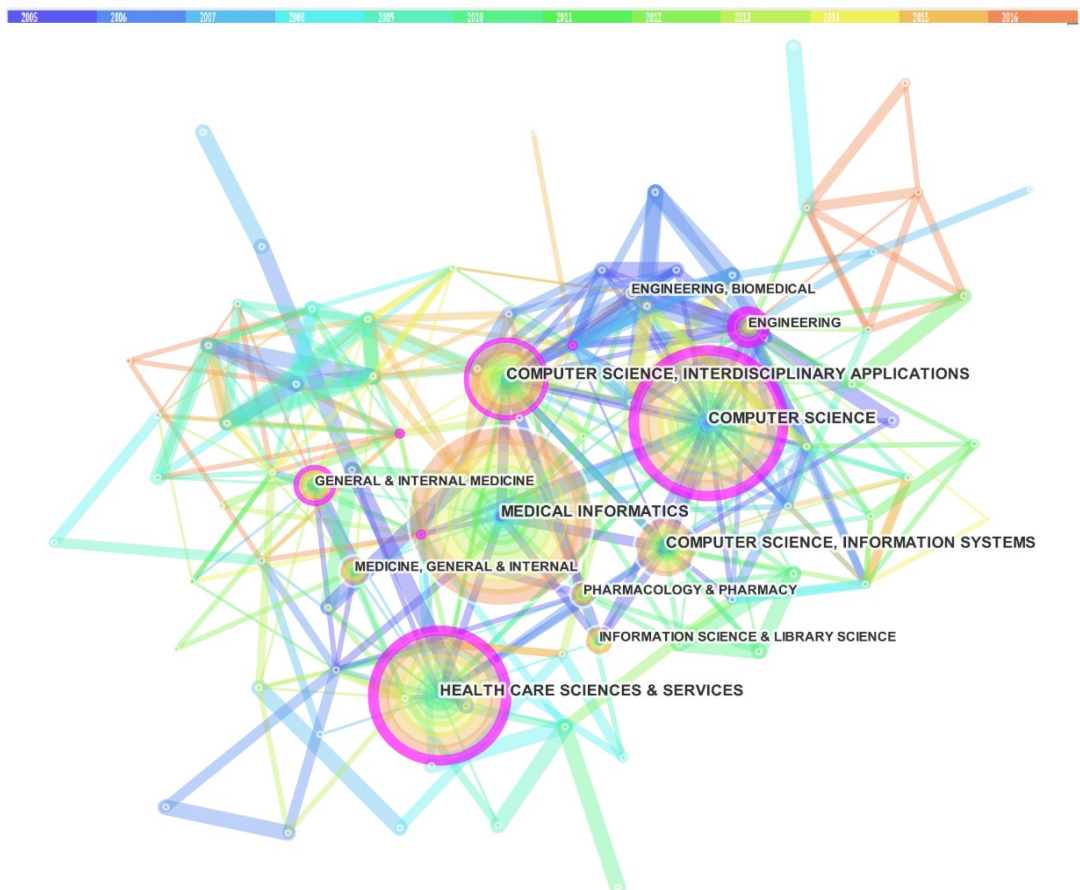

**Figure 10.** The category network containing *95* categories and *355* links. Concentric citation tree rings demonstrate the citation history of the publications of an institution. The purple circle represents betweenness centrality. The thicker the purple ring, the higher the centrality score. Medical Informatics is the category with highest frequency, whereas Health Care Sciences and Services is the most central category.

Table 15 lists the top *5* categories based on frequency. With the frequency of *658*, "Medical Informatics" leads the rest of the categories. Following it is the "Computer Science" with a frequency of *545*. Next is "Health Care Sciences & Services" with a frequency of *495*, which is followed by "Computer Science, Information Systems" and "Computer Science, Interdisciplinary Applications" with frequencies *320* and *318* respectively.

**Table 15.** The top 5 categories based on frequency. The subject category "Medical Informatics" leads over other categories with a frequency of 658.

| Frequency | Category |
|---|---|
| 658 | Medical Informatics |
| 545 | Computer Science |
| 495 | Health Care Sciences & Services |
| 320 | Computer Science, Information Systems |
| 318 | Computer Science, Interdisciplinary Applications |

After visually analysing co-authors, journals, co-cited authors, countries, institutions, and subject categories, in the end, we are presenting the summary of the results.

## 5. Summary of Results

In this paper, we have utilized CiteSpace for the analysis of various types of visualization to identify emerging trends and abrupt changes in scientific literature in the domain over time. In this section, we give an overview of the key results of the visual analysis performed in this study.

Firstly, using clustering of cited references we observed Cluster #*1*, the "computerised decision support" is the largest cluster, which contains *65* nodes that are *10.638%* of whole nodes in the network. The articles of Bates DW (*1999*), Stamos TD (*2001*), Van Der Sijs H (*2008*), and Middleton B (*2013*) are the key turning point. The half-life of these articles is *7, 4, 5, and 3* years respectively.

Subsequent analyses verified that there is conducted diversity in authors, journals, countries, institutions, and subject categories.

In the analysis of journals, we observed that the "Journal of the American Medical Informatics Association" has the largest number of highly cited publications in the domain and "Journal of the American Medical Association" is the most central journal among all other journals.

In terms of the analysis of the author's network, we observed that Ali S. Raja (*2014*) has the strongest burst among top all authors of the domain since *2005*. We also observed that most collaborative author in the network is David Bates, a Prof. of Medicine at the"Harvard School", has *59* citations is also the most central author with centrality score *0.33*. His areas of interest are medication safety, patient safety, quality, medical informatics, and clinical decision support. It is interesting to note that David Bates is also the highly cited and most central cited author of this domain.

In the analysis of countries, top *30* countries were chosen from the entire time span of *2005-2016* for each one-year time slice. We observed that the United States has the highest frequency, which indicates the origin of key publications in the domain. Whereas Canada has the highest centrality score. Scotland has the strongest citation burst, which provides the evidence that the articles originating in the domain from Scotland have attracted a degree of attention from its research community.

On the visual analysis of institutions, we found that "The University of Massachusetts" has the strongest and longest citation burst in the timespan of *2006* to *2009*. The "Indiana University School of Medicine" also has the longest period of the burst among all other institutes from *2013* till *2016*. Harvard has a top ranking with a frequency of *165* publications. It is interesting to note that the Harvard is also the most central institution with the centrality score *0.3*.

In the analysis of categories, we observed that the category "Health Care Sciences & Services" leads over other categories with centrality value *0.29*. Whereas with a frequency of *658*, the category "Medical Informatics" leads the rest of the categories.

## 6. Conclusions and Future Work

In this paper, we have demonstrated a comprehensive visual and scientometric survey of the CDSS domain. This research covers all Journal articles in Thomson Reuters from the period2005-2016. Our survey is based on real data from the Web of Science databases. This allowed us to comprehend all publications in the domain of CDSSs.

Our analysis has produced many interesting results. TheCDSS has gained the interest of the research community from the era of 2005. David Bates is the highly cited author in the literature of CDSS, whereas Ali S. Raja is the author who hasrapidly grown the number of publications during the period of study. The "Journal of the American Informatics Medical Association" is the top ranking source journal. It contributes 1169 publications during the period of study. The United States has contributed the highest number of publications, whereas the United Kingdom is the second highest productive country. Most of the contributions came from Harvard, whereas the "University of Massachusetts" remainedspecifically active in the research in this domain. The "HealthCare Sciences & Services" leads the rest of the categories inCDSS.

A significant dimension of future work is to conductscientometric analysis for identifying disease patterns,specifically in the cardiovascular, breast cancer and diabetesdomains

### Acknowledgment

This research project is funded by the EPSRC (Grant Ref. No. EP/H501584/1) and Sitekit Solutions Ltd. We would like to thank Professor Warner Slack from Harvard Medical School for providing useful insights and for his support and encouragement.